# Thermal Contribution in the Electrical Switching Experiments with Heavy Metal / Antiferromagnet Structures

Peng Sheng[†], Zhengyang Zhao[†], Onri Jay Benally, Delin Zhang, and Jian-Ping Wang[*]

Department of Electrical and Computer Engineering, University of Minnesota, 4-174 200 Union Street

SE, Minneapolis, MN 55455, USA

We examine the thermal origin of the detected "saw-tooth" shaped Hall resistance ($R_{xy}$) signals in the spin-orbit torque switching experiment for antiferromagnetic MnN. Compared with the results of the heavy metal / antiferromagnet bilayers (MnN/Ta), the qualitatively same "saw-tooth" shaped signals also appear in the samples with the heavy metal layer alone (either Ta or Pt) without MnN layer. In addition, The $R_{xy}$ signal changes oppositely in the devices with Ta and Pt, due to the opposite temperature coefficient of resistivity (TCR) of the two materials. All those results are consistent with the "localized Joule heating" mechanism in devices with Hall crosses geometry. Moreover, by utilizing a structure with separated writing current paths and Hall cross area, the quadratic relationship between $\Delta R_{xy}$ and the writing current's amplitude is observed, which provides quantitative evidence of the thermal contribution. These results reveal the dominant thermal artifact in the widely used Hall crosses geometry for Néel vector probing, and also provide a strategy to semi-quantitatively evaluate the thermal effect, which can shed light on a more conclusive experiment design.

[*]Corresponding author. Tel: (612) 625-9509. E-mail: jpwang@umn.edu

[†] These authors contributed equally to this work.

# I. INTRODUCTION

Antiferromagnets (AFMs) are magnetically ordered materials without net magnetization, which feature negligible magnetic stray fields, insensitivity against external magnetic field perturbations, and ultrafast spin dynamics [1–7], in contrast to ferromagnets. Thus, the prospects of developing reliable, high-speed, and high-density nonvolatile memory devices through antiferromagnetic materials are now recognized. [8,9] However, due to the absence of net magnetic moment, both the manipulation and detection of the antiferromagnetic states (Néel vector) are significantly challenging.

Various approaches have been explored to control the magnetic states in AFMs before, which usually rely on external magnetic fields [10–13], ferromagnets [14–17], optical excitation [18–23], terahertz pump pulse [24–29], strain and electric field [11,30–36] and so on [37]. Nevertheless, within the context of controllability and practical use, direct all-electrical switching and detection of the Néel vector are particularly desired and have attracted intensive interest in the antiferromagnetic spintronics study. Recently, breakthrough progresses have been made in metallic AFMs, such as CuMnAs and $Mn_2Au$ [38–41]. Their special crystal symmetries enable the generation of staggered Néel spin-orbit torques (NSOT) through applying electrical current, which efficiently switch the antiferromagnetic order on each opposite spin-sublattices. While the special requirement of broken inversion in crystallographic symmetries restricts the choice of potential antiferromagnetic materials. Most recently, spin Hall effect induced SOT switching of antiferromagnetic NiO has been reported, where NiO is interfaced to Pt [42–45]. A charge current in the Pt layer generates a spin current that can transport into NiO and switch its Néel orders without relying on the special crystal symmetries of AFMs and an external field. It could provide a more feasible route towards all-electrical manipulation in more AFMs with an adjacent heavy metal (HM) layer, such as α-$Fe_2O_3$/Pt and metallic MnN/Pt [46–48]. To electrically "read out" or detect the antiferromagnetic order switching, the measurement of Hall resistance signal ($R_{xy}$) is widely carried out on a Hall cross device with the 8-terminal geometry, as shown in Fig. 1(a). A "saw-tooth" [38–42,48] or "steplike" [44,46,47] shaped $R_{xy}$ signals are identified as the signature of antiferromagnetic order switching. However, recent studies found that the qualitatively same $R_{xy}$ signals also appeared in non-magnetic films, such as Pt and Nb in the absence of AFM layer, which suggested the crucial role of non-magnetic contribution in the electrical antiferromagnetic switching experiments with SOT [47,49–51]. The localized Joule heating and electromigration are considered as the origin of the probed $R_{xy}$ signals with "saw-tooth" and "steplike" shape respectively, since the high-density writing currents ($I_w$) is utilized [49–51]. As all the works mentioned above are mainly limited to samples consisting of antiferromagnetic insulator (AFI) and heavy metal Pt, it is essential to expand the study to other materials in order to examine and confirm the origin of the $R_{xy}$ signals.

In this paper, utilizing the 8-terminal geometry, we examine the origin of the "saw-tooth" shaped $R_{xy}$ signals observed in the electrical antiferromagnetic "switching" experiments with the MnN/Ta bilayers. MnN is a metallic AFM material with high Néel temperature [52]. Ta is a commonly used HM with *opposite* sign of SHA as Pt [53,54]. The same measurement is then repeated with two control samples without the MnN layer: Ta sample and Pt sample. The "saw-tooth" shaped $R_{xy}$ signals, which are similar to those observed in previous reports [38–42,46–50], is obtained in all three samples (MnN/Ta sample, Ta sample and Pt sample), indicating the non-magnetic (thermal) origin for such signals. In addition, the Ta sample and the Pt sample show *opposite* read out $R_{xy}$ signal, which is determined by the *opposite* temperature coefficient of resistivity (TCR) of the two HM, and not related to the *opposite* sign of the SHA between Ta and Pt. Moreover, by utilizing a geometry with separated writing and reading paths, the quadratic relationship between the "read out" signal $\Delta R_{xy}$ and the "writing" current is observed. This quadratic relationship can be explained through semi-quantitative analysis, which further manifests the decisive role of "localized Joule heating" in such AFM "switching" experiments. Owing to the thermal influence, the electrical detection method for AFM switching through Hall crosses geometry has to be re-examined and the thermal artifacts have to be carefully evaluated.

## II. SAMPLE PREPARATION AND METHODS

As stated above, we studied three film structures: the MnN/Ta sample [substrate/MnN(5)/Ta(10)], the Ta sample [substrate/Ta(7)/MgO(4)], and the Pt sample [substrate/Cr(2)/Pt(7)] (all thicknesses are in nm, and substrate here is thermally oxidized silicon substrate). All the films were grown using DC sputtering in a Shamrock sputter tool at room temperature (RT), except for the antiferromagnetic MnN thin films which is deposited in the same way as described in our previous work [55]. To confirm the antiferromagnetism of the MnN film, the magnetic hysteresis (M-H) loop of bilayers with MnN(25)/CoFeB(3) is measured. A distinct shift of 480 Oe in the hysteresis loop is observed compared to its coercivity of 250 Oe, as seen in Fig. 1(a), which manifests a strong exchange bias and verifies the antiferromagnetism of the MnN thin film.

The films were then patterned into 8-terminal Hall crosses geometry devices using photolithography and Ar ion milling procedures, as shown in Fig. 1(b). The width of the vertical and horizontal Hall terminals is 6 μm and the other four terminals (45° apart from the vertical and horizontal bars) are 10 μm wide. Then the metallic Hall crosses were covered by a 50 nm-thick $SiN_x$ film to avoid oxidation. For the measurement, a Keithley 6221 current source and a Keithley 2182A nanovolt meter were used to conduct the Hall measurement, and the other Keithley 6221 current source was used to generate the writing current ($I_w$). As depicted in Fig. 1(b), $I_w$ is injected through the writing paths along the +/− 45º directions. The Hall measurement is performed in the x and y direction, with a small probe current $I_{probe}$ injected along the x direction, and the Hall voltage $V_{Hall}$ measured along the y direction.

## III. RESULTS AND DISCUSSION

We first conducted the standard electrical SOT switching experiment procedures on the MnN(5)/Ta(10) sample: apply a sequence of writing current pulses and measure the Hall signal after each pulse. The pulse amplitude and pulse width are denoted in Fig. 1(c). Firstly, five current pulses are applied along the −45º axis [left-hand (LH) writing path] and $R_{xy}$ after each pulse is gradually increased (from negative to positive value). Then, five pulses with the same amplitude and width are applied along the +45º axis [right-hand (RH) writing path], and $R_{xy}$ decreases accordingly (from positive to negative value). Reversing the direction of the pulse by 180º won't affect such trend on $R_{xy}$. We repeated this experimental procedure for several cycles, and Fig. 1(c) shows the change of $R_{xy}$ within one cycle. Such "saw-tooth" shaped signal is consistent with the AFM/Pt system as reported before [42,46–50]

To clarify the mechanism of the probed "saw-tooth" shaped $R_{xy}$ signal in MnN/Ta sample, we performed the same experiment on the two control samples: the Ta sample and the Pt sample without MnN layer. Both of the control samples are patterned in the same way as shown in Fig. 1(b). The same "switching" experiment procedures were carried out on them: 10 LH writing pulses followed by 10 RH writing pulses, with $R_{xy}$ measured after each pulse. The recorded $R_{xy}$ signals of the Ta and Pt samples are plotted in Fig. 2(a) and 2(b), respectively. Surprisingly, both figures show the "saw-tooth" shaped $R_{xy}$ signals even though there is no antiferromagnetic MnN layer existent. These results of the control samples reflect the non-magnetic origin of the "saw-tooth" shaped $R_{xy}$ signals. In addition, it is interesting to find $R_{xy}$ for the Pt sample changes in the *opposite* direction to that of the Ta and MnN/Ta samples [by comparing Fig. 2(b) with Fig. 2(a) and Fig. 1(c)]. This phenomenon is very easily associated with the *opposite* sign of SHA in Ta and Pt [53,54]. However, this possibility can be excluded because: (a) the change of $R_{xy}$ is obtained even without any FM or AFM layer, indicating the non-magnetic origin of the $R_{xy}$ signal, and (b) even if for AFM/HM bilayers, and assuming the change of $R_{xy}$ is mainly determined by the Néel vector of AFM which is controllable by SOT, the polarity of $R_{xy}$ still wouldn't be affected by the sign of SHA, because $R_{xy}$ will have 2-fold symmetry as the the Néel vector rotates from 0º to 360º.

Now we infer that the thermal artifact could be the origin of the "saw-tooth" shaped $R_{xy}$ signal, which is also pointed out in Refs. [50,51]. Upon the application of a writing current (along the +/− 45º path), the whole writing path is heated up by the electrical current. Especially, within the Hall cross area at the center of the device, the heating is non-uniform due to the inhomogeneous distribution of writing current density.

Higher temperature appears around the two constricting corners ("hot spots"), as illustrated in Fig. 2(c). The resistivity of the film will be changed with the temperature. Then, during the Hall measurement stages, the path of $I_{probe}$ is deviated due to the inhomogeneous resistivity. The deviation of $I_{probe}$ then leads to the variation of the $R_{xy}$ signal. When the writing path is changed from LH to RH, the deviation of $I_{probe}$ is also changed, and the trend of $R_{xy}$ signal reversed accordingly. In short, it is the inhomogeneous heating of $I_w$ in the Hall cross area that generates the "saw-tooth" shaped $R_{xy}$ signal.

In addition, we still need to figure out in detail how the resistivity changes with temperature, and why the $R_{xy}$ signal of the Ta sample and the Pt sample is *opposite*. Therefore, we measured the resistivity of both Ta and Pt films as a function of temperature, which is plotted in Fig. 2(d). We found that the resistivity of Ta film decreases, whereas the resistivity of Pt film increases as the increase of the temperature, revealing the *opposite* signs of TCR for Ta and Pt. It should be noted that the negative sign of TCR for Ta reflects the formation of tetragonal β phase [56]. Now, the *opposite* trends of $R_{xy}$ for Ta and Pt can be well explained. When applying $I_w$, the thermal effect and the *opposite* signs of TCR induced *opposite* changes in the resistivity of Ta and Pt. Then the path of $I_{probe}$ is deviated in *opposite* ways within the Ta film and the Pt film, resulting in the *opposite* $R_{xy}$ responses, as illustrated in Figs. 2(e) and 2(f). From the analysis above, it can thus be concluded that all of the experimental results are in accordance with the "localized Joule heating" mechanism [50,51].

To get a better understanding of thermal influence on the detected $R_{xy}$ signals, we designed a second schematic, where the "writing" or "heating" paths are electrically isolated from the Hall measurement cross, as illustrated in Fig. 3(a). First, the HM film (either Ta or Pt) is patterned into a 4-terminal cross bar along the x and y direction (for the Hall measurement). Then, a 50-nm thick $SiN_x$ film is deposited on the HM cross bar for electrical isolation. Finally, another cross along the +/− 45º directions is deposited, which consists of Ti (60)/Au (60) films. The top cross bar is used as the heating source. In this way, we can simultaneously apply the heating/writing current on the top paths and track the $R_{xy}$ signals on the bottom

HM layer. Consequently, the detected $R_{xy}$ signal directly reflects the influence of Joule heating. Both the Ta and the Pt films were patterned into this schematic, and the measurement results are shown in Figs. 3(b) and 3(c), respectively.

The experiment procedure is described as follows: initially, a DC current $I$ with a fixed amplitude is applied along −45º axis (LH path) for 600 s. Then $I$ is off for 600 s to cool the device down to room temperature. After that, $I$ is applied along +45º axis (RH path) for 600 s, and then turned off for cooling down. Such an experiment procedure is repeated for several cycles. The $R_{xy}$ signal is simultaneously recorded throughout this procedure. As depicted in Figs. 3(b) and 3(c), the $R_{xy}$ signals show a periodic change depending on the direction of $I$: for LH heating, $R_{xy}$ of Ta increases and $R_{xy}$ of Pt decreases; for RH heating, $R_{xy}$ of Ta decreases and $R_{xy}$ of Pt increases; with the heating current off, $R_{xy}$ gradually returns to 0. These trends are in agreement with the result of the 8-terminal devices shown in Figs. 2(a) and 2(b). Again, the *opposite* incremental change of the $R_{xy}$ signal for the Ta sample and the Pt sample confirms the critical role of Joule heating influenced HM layer's resistive property in the Hall measurement.

The $R_{xy}$ signal in Fig. 3 is in "steplike" shape, because the Joule heating produced by a continuous DC current is more remarkable than that of pulse current. When the DC current is turned on, the temperature of the Hall cross area (in the HM layer) rises rapidly to a saturated value, so does $R_{xy}$. Similarly, when the DC current is turned off, the temperature drops rapidly from the saturated point, and $R_{xy}$ returns to 0. It can also be seen the $R_{xy}$ signal of the Pt sample responds much faster than the Ta sample to the heating current. This could be due to the higher thermal conductivity of Pt, which makes the Pt sample reach to thermal equilibrium much faster than the Ta sample. It is noteworthy that, $R_{xy}$ is not originated from the *Seebeck* effect (as opposed to the explanation in [49]), since the *Seebeck* voltage is negligible compared to the signal obtained here (as discussed in the Supplementary Material). The variation of $R_{xy}$ is resulted from the deviation of $I_{probe}$ due to the "localized Joule heating" mechanism.

It is also noticed that in Figs. 3(b) and 3(c), the amplitude of $R_{xy}$ increases with the increase of the DC

current. In order to have a quantitative analysis on the relationship between the Joule heating and the $R_{xy}$ signal, $\Delta R_{xy}$ [the amplitude of $R_{xy}$ in Figs. 3(b) and 3(c)] is plotted as a function of the heating current $I$ in Figs. 4(a) and 4(b), for the Ta and Pt samples respectively. From the fitting curves, we can see $\Delta R_{xy}$ and the heating current $I$ follows quadratic relationship. This quadratic relationship (expressed as $\Delta R_{xy} = \phi \cdot I^2$, where $\phi$ is a constant) can be derived from the *Joule effect* and the detailed derivation steps can be found in the Appendix. This result furtherly confirms the decisive influence of Joule heating on the generation of the $R_{xy}$ signal.

## IV. CONCLUSIONS

In summary, the significant role of "localized Joule heating" in the electrical SOT switching experiment for MnN/Ta was systematically examined. We confirmed that the "saw-tooth" shaped $R_{xy}$ signals commonly appeared in the devices with 8-terminal Hall crosses geometry, even without an antiferromagnetic layer in the film stack. In addition, the opposite $R_{xy}$ signals for the Ta sample and Pt sample were observed, which were attributed to the opposite signs of temperature coefficient of resistivity in Ta and Pt. These results are fully compatible with the "localized Joule heating" mechanism, which is the artifact of the high-density writing current. In addition, in the device with the writing paths layer and the HM layer electrically isolated, the $\Delta R_{xy}$ changes quadratically as the function of DC writing current's amplitude. It provides a quantitative evidence that the "localized Joule heating" mechanism dominantly influences the generation of the $R_{xy}$ signals and demonstrates that the thermal properties of HM layer carry out a critical role in the "read out" $R_{xy}$ signals. We thus consider that the electrical detection of the Néel vector state through SOT switching experiment is not conclusive without fully examining the thermal influence of the writing current. Our study solved the puzzle of Joule heating's influence and HM layer's role in the "read out" $R_{xy}$ signals and provides insights in the electrical SOT switching experiments for AFMs.

All the experiments presented in this paper were carried out and completed before Sep. 2018.

**SUPPLEMENTARY MATERIAL**

See Supplementary Material for extra experimental results, including the $R_{xy}$ signals with various writing pulse widths and pulse amplitudes, the decay of $R_{xy}$ signal after shutting off the writing current and the measurement of the *Seebeck* voltage.

**CONTRIBUTIONS**

J.-P.W. initiated the study and coordinated the overall project. Z.Z. conceived the experiments and performed the electrical measurements. Z.Z. and O.J.B carried out the device fabrication. D.Z. prepared the MnN film and characterized the exchange bias of MnN. P.S. and Z.Z. consolidated the analysis and interpretation of the data. P.S., Z.Z. and J.-P.W discussed, outlined and drafted the manuscript. All authors discussed the results and commented on the manuscript.

**ACKNOWLEDGMENTS**

This work was initially supported by C-SPIN, one of six centers of STARnet, a Semiconductor Research Corporation program, sponsored by MARCO and DARPA. This work was later supported in part by ASCENT, one of six centers in JUMP, a Semiconductor Research Corporation (SRC) program sponsored by DARPA. Portions of this work were conducted in the Minnesota Nano Center, which is supported by the National Science Foundation through the National Nano Coordinated Infrastructure Network (NNCI) under Award Number ECCS-2025124.

# APPENDIX: DERIVATION OF THE QUADRATIC RELATIONSHIP BETWEEN $\Delta R_{xy}$ AND $I$

When the temperature is not very high and varies in a moderate range, the metallic material's resistivity approximately changes linearly as the function of temperature [this is also illustrated in Fig. 2(d)). It can be expressed as:

$$\rho_T = \rho_0(1+\alpha T), \qquad (1)$$

where $\rho_T$ and $\rho_0$ is the resistivity of metallic material at temperatures $T$ and $T_0$, respectively. $\alpha$ represents the TCR of the metal, which is a constant. As we discussed above, Ta and Cr/Pt possess opposite sign of $\alpha$, see Fig. 2(d). According to equation (1), the variation of resistance ($\Delta R$) due to the change of temperature $\Delta T = T - T_0$ is:

$$\Delta R = \beta \bullet \rho_0 \bullet \alpha \Delta T, \qquad (2)$$

where $\beta$ is the geometry factor of the material.

According to the *Joule effect*, the Joule heat ($\Delta Q$) generated by DC current ($I$) in the conductive current path [the writing path as shown in Fig. 3(a), which consists of Ti/Au bilayers] is expressed as:

$$\Delta Q = I^2 \bullet R_w \bullet t, \qquad (3)$$

where $R_w$ represents the resistance of the Ti/Au writing path and $t$ represents the length of time the DC current passed through. Besides, when heat energy ($\Delta Q'$) is added to a substance, the temperature will change, their relationship is described as:

$$\Delta Q' = c \bullet m \bullet \Delta T, \qquad (4)$$

where $c$ and $m$ are the specific heat capacity and the mass of the material, respectively. If the Joule heat produced from the Ti/Au writing path transfers to the Hall cross HM layer with heat transfer efficiency $\varepsilon$, then $\Delta Q' = \varepsilon \bullet \Delta Q$. Therefore, according to equations (3) and (4),

$$c \cdot m \cdot \Delta T = \varepsilon \cdot I^2 \cdot R_w \cdot t, \tag{5}$$

thus,
$$\Delta T = \frac{\varepsilon \cdot I^2 \cdot R_w \cdot t}{c \cdot m}. \tag{6}$$

By combining equation (6) with equation (2),

$$\Delta R = \beta \cdot \rho_0 \cdot \alpha \Delta T = \frac{\beta \cdot \rho_0 \cdot \alpha \cdot \varepsilon \cdot I^2 \cdot R_w \cdot t}{c \cdot m} = \chi \cdot I^2, \tag{7}$$

where $\chi = \frac{\beta \cdot \rho_0 \cdot \alpha \cdot \varepsilon \cdot R_w \cdot t}{c \cdot m}$ is a constant for a sample. Due to the deviation of $I_{probe}$, the detected $R_{xy}$ signal actually is a portion of the $R$ (which is the $R_{xx}$ signal). Thus,

$$\Delta R_{xy} = \varphi \cdot \Delta R = \varphi \cdot \chi \cdot I^2 = \phi \cdot I^2, \tag{8}$$

where $\varphi$ is a constant which characterizes the portion of the $R_{xx}$ signal with the value between 0 and 1. Equation (8) coincides well with the fitting result plotted in Fig. 4. In addition, the sign of $\Delta R_{xy}$ determined by the TCR coefficient $\alpha$ is also well described in equations (7) and (8).

**REFERENCES**

[1] A. N. Bogdanov, A. V. Zhuravlev, and U. K. Rößler, Phys. Rev. B **75**, 094425 (2007).

[2] O. Gomonay, T. Jungwirth, and J. Sinova, Phys. Rev. Lett. **117**, 017202 (2016).

[3] P. E. Roy, R. M. Otxoa, and J. Wunderlich, Phys. Rev. B **94**, 014439 (2016).

[4] T. Jungwirth, X. Marti, P. Wadley, and J. Wunderlich, Nat. Nanotechnol. **11**, 231 (2016).

[5] K. Olejník, T. Seifert, Z. Kašpar, V. Novák, P. Wadley, R. P. Campion, M. Baumgartner, P. Gambardella, P. Nemec, J. Wunderlich, J. Sinova, P. Kužel, M. Müller, T. Kampfrath, and T. Jungwirth, Sci. Adv. **4**, eaar3566 (2018).

[6] P. Bowlan, S. A. Trugman, D. A. Yarotski, A. J. Taylor, and R. P. Prasankumar, J. Phys. D. Appl. Phys. **51**, 194003 (2018).

[7] V. Baltz, A. Manchon, M. Tsoi, T. Moriyama, T. Ono, and Y. Tserkovnyak, Rev. Mod. Phys. **90**, 015005 (2018).

[8] H. Seinige, M. Williamson, S. Shen, C. Wang, G. Cao, J. Zhou, J. B. Goodenough, and M. Tsoi, Phys. Rev. B **94**, 214434 (2016).

[9] J. Wang, S. S. Sapatnekar, C. H. Kim, P. Crowell, S. Koester, S. Datta, K. Roy, A. Raghunathan, X. S. Hu, M. Niemier, A. Naeemi, C.-L. Chien, C. Ross, and R. Kawakami, in *Proc. 54th Annu. Des. Autom. Conf. 2017* (ACM, New York, NY, USA, 2017), pp. 1–6.

[10] G. R. Hoogeboom, A. Aqeel, T. Kuschel, T. T. M. Palstra, and B. J. van Wees, Appl. Phys. Lett. **111**, 052409 (2017).

[11] A. A. Sapozhnik, R. Abrudan, Y. Skourski, M. Jourdan, H. Zabel, M. Kläui, and H.-J. Elmers, Phys. Status Solidi - Rapid Res. Lett. **11**, 1600438 (2017).

[12] X. Marti, I. Fina, C. Frontera, J. Liu, P. Wadley, Q. He, R. J. Paull, J. D. Clarkson, J. Kudrnovský, I. Turek, J. Kuneš, D. Yi, J. H. Chu, C. T. Nelson, L. You, E. Arenholz, S. Salahuddin, J. Fontcuberta, T. Jungwirth, and R. Ramesh, Nat. Mater. **13**, 367 (2014).

[13] D. Petti, E. Albisetti, H. Reichlová, J. Gazquez, M. Varela, M. Molina-Ruiz, A. F. Lopeandía, K. Olejník, V. Novák, I. Fina, B. Dkhil, J. Hayakawa, X. Marti, J. Wunderlich, T. Jungwirth, and R. Bertacco, Appl. Phys. Lett. **102**, 192404 (2013).

[14] B. G. Park, J. Wunderlich, X. Martí, V. Holý, Y. Kurosaki, M. Yamada, H. Yamamoto, A. Nishide, J. Hayakawa, H. Takahashi, A. B. Shick, and T. Jungwirth, Nat. Mater. **10**, 347 (2011).

[15] Y. Y. Wang, C. Song, B. Cui, G. Y. Wang, F. Zeng, and F. Pan, Phys. Rev. Lett. **109**, 137201 (2012).

[16] A. Scholl, M. Liberati, E. Arenholz, H. Ohldag, and J. Stöhr, Phys. Rev. Lett. **92**, 247201 (2004).

[17] X. Martí, B. G. Park, J. Wunderlich, H. Reichlová, Y. Kurosaki, M. Yamada, H. Yamamoto, A. Nishide, J. Hayakawa, H. Takahashi, and T. Jungwirth, Phys. Rev. Lett. **108**, 017201 (2012).

[18] A. V. Kimel, A. Kirilyuk, A. Tsvetkov, R. V. Pisarev, and T. Rasing, Nature **429**, 850 (2004).

[19] A. V. Kimel, C. D. Stanciu, P. A. Usachev, R. V. Pisarev, V. N. Gridnev, A. Kirilyuk, and T. Rasing, Phys. Rev. B **74**, 060403 (2006).

[20] R. V. Mikhaylovskiy, E. Hendry, V. V. Kruglyak, R. V. Pisarev, T. Rasing, and A. V. Kimel, Phys. Rev. B **90**, 184405 (2014).

[21] M. Fiebig, K. Miyano, Y. Tokura, and Y. Tomioka, Science **280**, 1925 (1998).

[22] D. Afanasiev, B. A. Ivanov, A. Kirilyuk, T. Rasing, R. V. Pisarev, and A. V. Kimel, Phys. Rev. Lett. **116**, 097401 (2016).

[23] A. V. Kimel, B. A. Ivanov, R. V. Pisarev, P. A. Usachev, A. Kirilyuk, and T. Rasing, Nat. Phys. **5**, 727 (2009).

[24] T. Kampfrath, A. Sell, G. Klatt, A. Pashkin, S. Mährlein, T. Dekorsy, M. Wolf, M. Fiebig, A. Leitenstorfer, and R. Huber, Nat. Photonics **5**, 31 (2011).

[25] S. Baierl, M. Hohenleutner, T. Kampfrath, A. K. Zvezdin, A. V. Kimel, R. Huber, and R. V. Mikhaylovskiy, Nat. Photonics **10**, 715 (2016).

[26] Y. Mukai, H. Hirori, T. Yamamoto, H. Kageyama, and K. Tanaka, New J. Phys. **18**, 013045 (2016).

[27] S. Baierl, J. H. Mentink, M. Hohenleutner, L. Braun, T. M. Do, C. Lange, A. Sell, M. Fiebig, G. Woltersdorf, T. Kampfrath, and R. Huber, Phys. Rev. Lett. **117**, 197201 (2016).

[28] T. Kubacka, J. A. Johnson, M. C. Hoffmann, C. Vicario, S. De Jong, P. Beaud, S. Grübel, S. W. Huang, L. Huber, L. Patthey, Y. D. Chuang, J. J. Turner, G. L. Dakovski, W. S. Lee, M. P. Minitti, W. Schlotter, R. G. Moore, C. P. Hauri, S. M. Koohpayeh, V. Scagnoli, G. Ingold, S. L. Johnson, and U. Staub, Science **343**, 1333 (2014).

[29] T. Satoh, R. Iida, T. Higuchi, M. Fiebig, and T. Shimura, Nat. Photonics **9**, 25 (2015).

[30] C. Bordel, J. Juraszek, D. W. Cooke, C. Baldasseroni, S. Mankovsky, J. Minár, H. Ebert, S. Moyerman, E. E. Fullerton, and F. Hellman, Phys. Rev. Lett. **109**, 117201 (2012).

[31] R. O. Cherifi, V. Ivanovskaya, L. C. Phillips, A. Zobelli, I. C. Infante, E. Jacquet, V. Garcia, S. Fusil, P. R. Briddon, N. Guiblin, A. Mougin, A. A. Ünal, F. Kronast, S. Valencia, B. Dkhil, A. Barthélémy, and M. Bibes, Nat. Mater. **13**, 345 (2014).

[32] T. Zhao, A. Scholl, F. Zavaliche, K. Lee, M. Barry, A. Doran, M. P. Cruz, Y. H. Chu, C. Ederer, N. A. Spaldin, R. R. Das, D. M. Kim, S. H. Baek, C. B. Eom, and R. Ramesh, Nat. Mater. **5**, 823 (2006).

[33] D. Lebeugle, D. Colson, A. Forget, M. Viret, A. M. Bataille, and A. Gukasov, Phys. Rev. Lett. **100**, 227602 (2008).

[34] T. Kosub, M. Kopte, R. Hühne, P. Appel, B. Shields, P. Maletinsky, R. Hübner, M. O. Liedke, J. Fassbender, O. G. Schmidt, and D. Makarov, Nat. Commun. **8**, 13985 (2017).

[35] X. He, Y. Wang, N. Wu, A. N. Caruso, E. Vescovo, K. D. Belashchenko, P. A. Dowben, and C. Binek, Nat. Mater. **9**, 579 (2010).

[36] Z. Zhao, W. Echtenkamp, M. Street, C. Binek, and J. P. Wang, in *Device Res. Conf. - Conf. Dig.*

*DRC* (IEEE, 2016), pp. 1–2.

[37] C. Song, Y. You, X. Chen, X. Zhou, Y. Wang, and F. Pan, Nanotechnology **29**, 112001 (2018).

[38] P. Wadley, B. Howells, J. Elezny, C. Andrews, V. Hills, R. P. Campion, V. Novak, K. Olejnik, F. Maccherozzi, S. S. Dhesi, S. Y. Martin, T. Wagner, J. Wunderlich, F. Freimuth, Y. Mokrousov, J. Kune, J. S. Chauhan, M. J. Grzybowski, A. W. Rushforth, K. W. Edmonds, B. L. Gallagher, and T. Jungwirth, Science **351**, 587 (2016).

[39] M. Meinert, D. Graulich, and T. Matalla-Wagner, Phys. Rev. Appl. **9**, 064040 (2018).

[40] S. Y. Bodnar, L. Šmejkal, I. Turek, T. Jungwirth, O. Gomonay, J. Sinova, A. A. Sapozhnik, H.-J. Elmers, M. Kläui, and M. Jourdan, Nat. Commun. **9**, 348 (2018).

[41] X. F. Zhou, J. Zhang, F. Li, X. Z. Chen, G. Y. Shi, Y. Z. Tan, Y. D. Gu, M. S. Saleem, H. Q. Wu, F. Pan, and C. Song, Phys. Rev. Appl. **9**, 054028 (2018).

[42] X. Z. Chen, R. Zarzuela, J. Zhang, C. Song, X. F. Zhou, G. Y. Shi, F. Li, H. A. Zhou, W. J. Jiang, F. Pan, and Y. Tserkovnyak, Phys. Rev. Lett. **120**, 207204 (2018).

[43] T. Moriyama, K. Oda, T. Ohkochi, M. Kimata, and T. Ono, Sci. Rep. **8**, 14167 (2018).

[44] L. Baldrati, O. Gomonay, A. Ross, M. Filianina, R. Lebrun, R. Ramos, C. Leveille, F. Fuhrmann, T. R. Forrest, F. MacCherozzi, S. Valencia, F. Kronast, E. Saitoh, J. Sinova, and M. Klaüi, Phys. Rev. Lett. **123**, 177201 (2019).

[45] I. Gray, T. Moriyama, N. Sivadas, G. M. Stiehl, J. T. Heron, R. Need, B. J. Kirby, D. H. Low, K. C. Nowack, D. G. Schlom, D. C. Ralph, T. Ono, and G. D. Fuchs, Phys. Rev. X **9**, 041016 (2019).

[46] P. Zhang, J. Finley, T. Safi, and L. Liu, Phys. Rev. Lett. **123**, 247206 (2019).

[47] Y. Cheng, S. Yu, M. Zhu, J. Hwang, and F. Yang, Phys. Rev. Lett. **124**, 027202 (2020).

[48] M. Dunz, T. Matalla-Wagner, and M. Meinert, Phys. Rev. Res. **2**, 013347 (2020).

[49] C. C. Chiang, S. Y. Huang, D. Qu, P. H. Wu, and C. L. Chien, Phys. Rev. Lett. **123**, 227203 (2019).

[50] A. Churikova, D. Bono, B. Neltner, A. Wittmann, L. Scipioni, A. Shepard, T. Newhouse-Illige, J. Greer, and G. S. D. Beach, Appl. Phys. Lett. **116**, 022410 (2020).

[51] T. Matalla-Wagner, J.-M. Schmalhorst, G. Reiss, N. Tamura, and M. Meinert, Phys. Rev. Res. **2**, 033077 (2020).

[52] M. Meinert, B. Büker, D. Graulich, and M. Dunz, Phys. Rev. B **92**, 144408 (2015).

[53] L. Liu, C.-F. Pai, Y. Li, H. W. Tseng, D. C. Ralph, and R. a Buhrman, Science **336**, 555 (2012).

[54] S. Emori, U. Bauer, S.-M. Ahn, E. Martinez, and G. S. D. Beach, Nat. Mater. **12**, 611 (2013).

[55] J. Liu, D. Zhang, K. Wu, X. Hang, and J. P. Wang, J. Phys. D. Appl. Phys. **53**, 035003 (2020).

[56] N. Schwartz, W. A. Reed, P. Polash, and M. H. Read, Thin Solid Films **14**, 333 (1972).

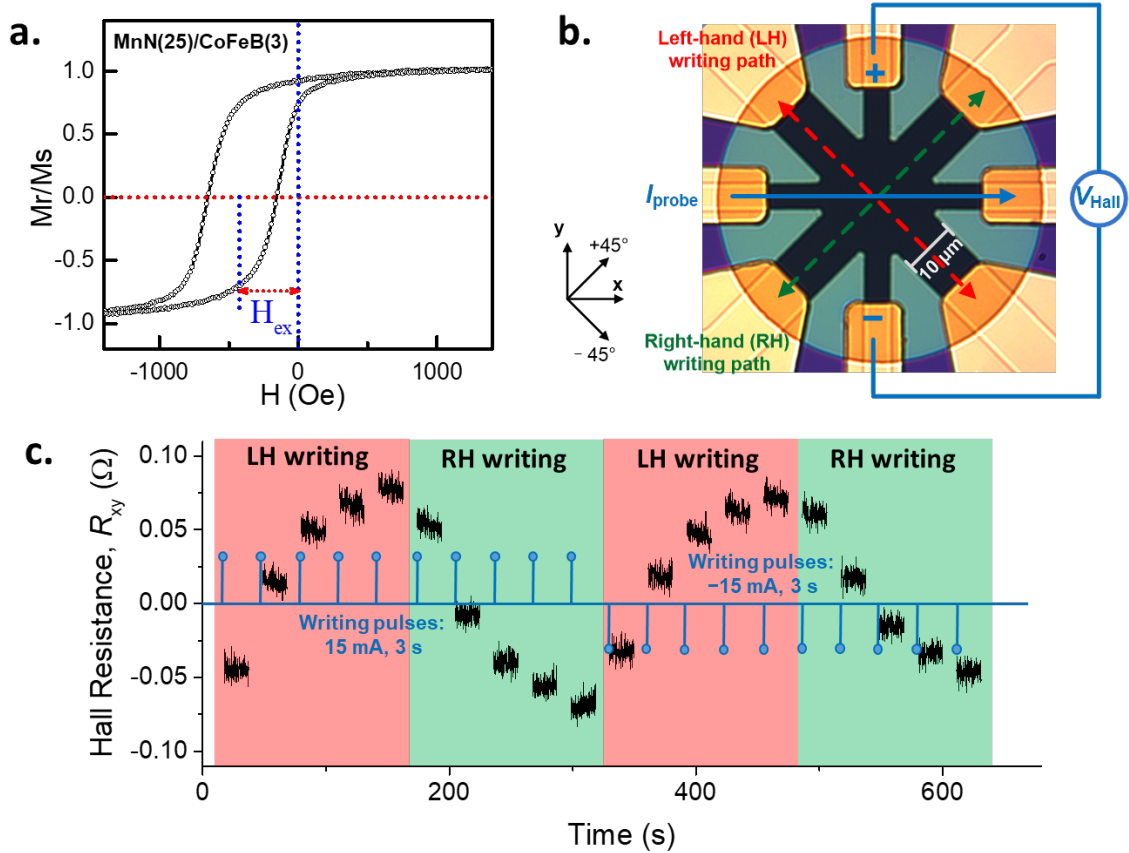

FIG. 1. (a) M-H loop of MnN(25)/CoFeB(3) bilayers showing the exchange bias with AFM MnN. (b) Image of the patterned 8-terminal Hall device. Two writing paths are along the +/− 45º directions (denoted as RH/LH writing path). And the Hall resistance ($R_{xy}$) can be measured by applying a small probe current $I_{probe}$ along the x direction and detecting the Hall voltage $V_{Hall}$ along the y direction. (c) The $R_{xy}$ signal of MnN(5)/Ta(10) sample obtained by conducting the standard electrical SOT switching experiment procedures. A sequence of writing pulses is applied, and the $R_{xy}$ is recorded after each pulse (with $I_{probe}$ = 0.1 mA). The writing pulses are either injected through the LH path (for the regions with red background), or through the RH path (for the regions with green background). The pulse amplitudes and pulse widths are denoted on the figure (pulses are denoted by the blue sticks). The $R_{xy}$ signal (black) shows "saw-tooth" shape.

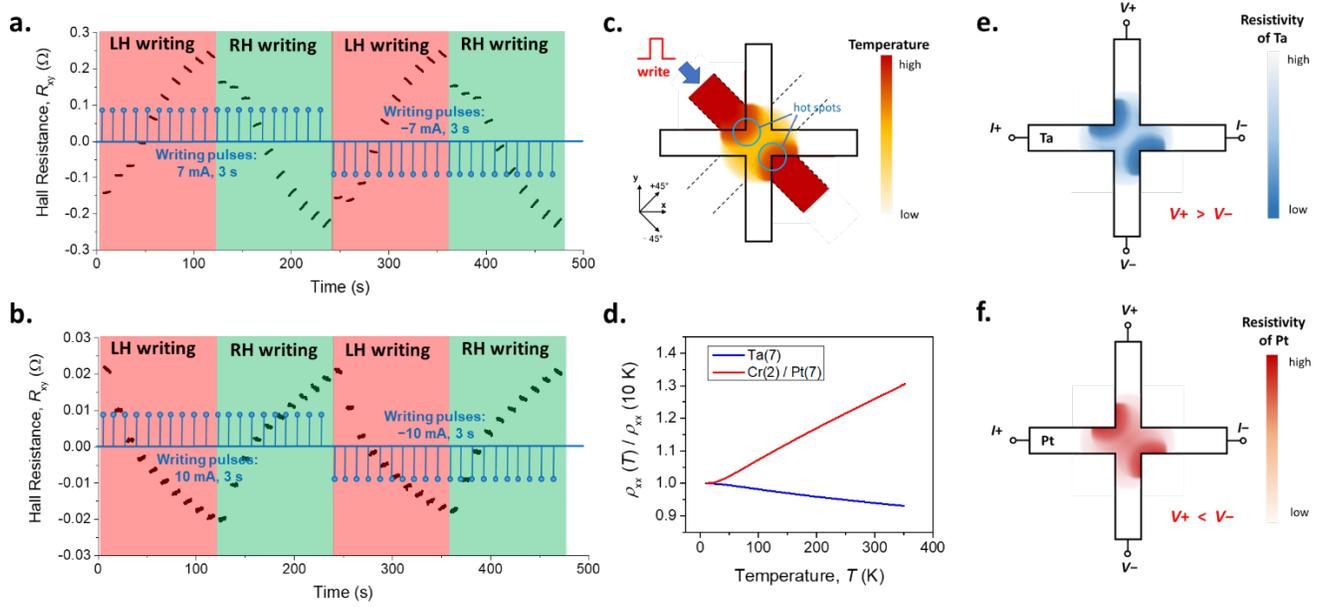

FIG. 2. The "saw-tooth" shaped $R_{xy}$ signal obtained from (a) the Ta sample, and (b) the Pt sample without MnN layer. $I_{probe} = 0.1$ mA. (c) Illustration of "localized Joule heating" when the writing current is injected through the LH writing path. (d) Normalized resistivity of the Ta sample (blue) and the Pt sample (red) as a function of temperature. (e) Illustration of the inhomogeneous distribution of resistivity for the Ta sample after the injection of LH writing pulse. The resistivity gets smaller around the "hot spots". (f) Illustration of the inhomogeneous distribution of resistivity for the Pt sample after the injection of LH writing pulse. The resistivity gets larger around the "hot spots".

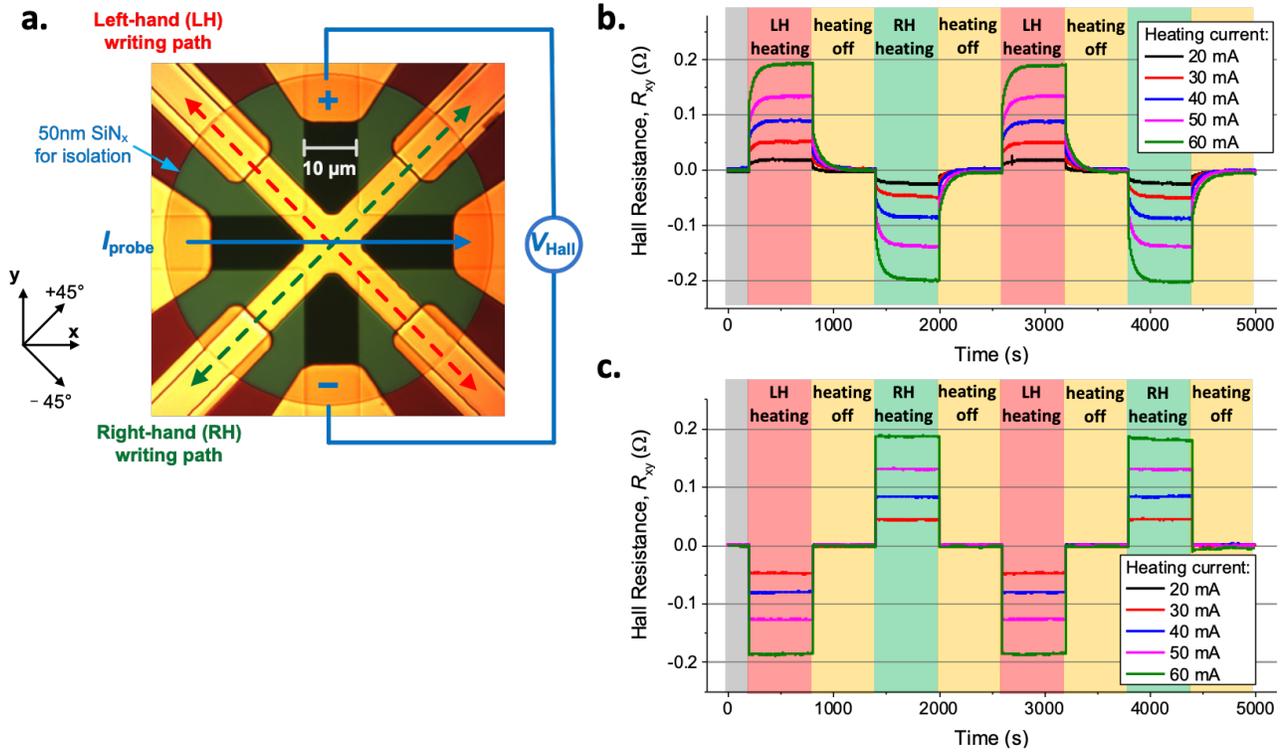

FIG. 3. (a) Image of the second schematic, where the writing (heating) paths are electrically isolated from the Hall measurement cross (HM layer). The HM Hall cross (either Ta or Pt) is first covered by a $SiN_x$ layer. Then the writing paths (along the +/− 45º directions) composed of Ti (60) / Au (60) are deposited on top of the $SiN_x$ layer. (b) and (c): $R_{xy}$ signals of (b) the Ta sample and (c) the Pt sample with the application of DC heating current on the writing paths. The heating current is either injected through the LH path (for the regions with red background), or through the RH path (for the regions with green background), or turned off (for the regions with yellow background). Heating current of different amplitudes are applied. $I_{probe}$ = 0.1 mA.

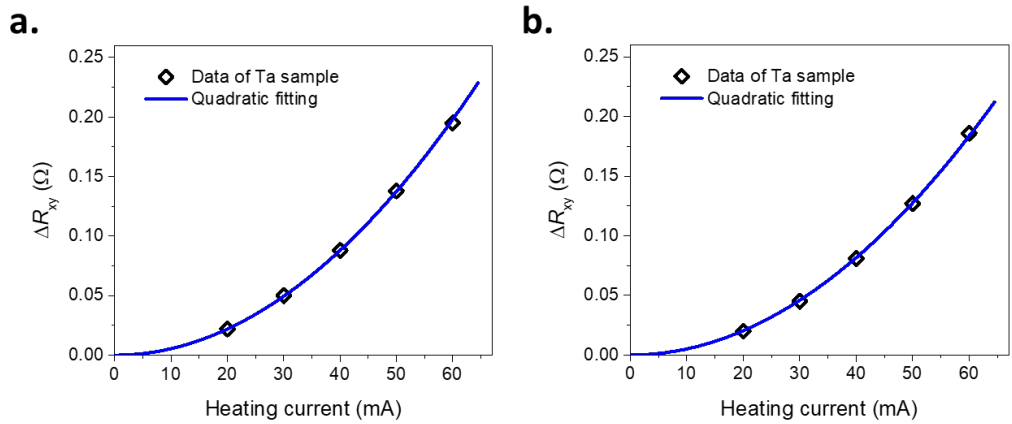

FIG. 4. The amplitude of the $R_{xy}$ signal changes as a function of heating current amplitude for (a) the Ta sample and (b) the Pt sample, extracted from Fig. 3(b) and 3(c), respectively. The blue lines are quadratic fittings.